\documentclass[10pt,final,journal]{IEEEtran}

\usepackage{graphicx}
\usepackage{cite}

\begin{document}

\title{Performance improvement of refractometric sensors through  hybrid plasmonic-Fano resonances}

\author{Mahmoud H. Elshorbagy,
 Alexander Cuadrado,
  Gabriel Gonz\'{a}lez, \\
   Francisco J. Gonz\'{a}lez~\IEEEmembership{Senior Member,~IEEE}, 
     and Javier Alda~\IEEEmembership{Member,~OSA}}%

\thanks{ M. H. Elshorbagy, A. Cuadrado, and J. Alda are with  Applied Optics Complutense Group. University Complutense of Madrid. Faculty of Optics and Optometry. Av. Arcos de Jalon, 118. 28037 Madrid, Spain.}%
\thanks{ M. H. Elshorbagy is also with Physics Department. Faculty of Science. Minia University.  61519 El-Minya, Egypt.} %
\thanks{ G. Gonz\'{a}lez, and F. J. Gonz\'{a}lez are with Universidad Aut\'{o}noma de San Luis Potos\'{\i}. Av. Sierra Leona, 550, Col. Lomas 2a. Secci\'{o}n. 78210 
San Luis Potos\'{\i}, M\'{e}xico.} %
\thanks{Corresponding author: Javier Alda, javier.alda@ucm.es}


\maketitle


\begin{abstract}

In this paper, we present  a plasmonic refractometric sensor that works under normal incidence; allowing its integration on a fiber tip. 
The sensor's material and geometry exploit the large scattering cross-section given by high-contrast of the index of refraction subwavelength dielectric gratings.
Our design generates a hybrid plasmonic-Fano resonance due to the interference between the surface plasmon resonance and the grating response. 
We optimize the sensor with a merit function that combines the quality parameter of the resonance and the field enhancement at the interaction volume where the plasmon propagates. Our device shows a  high sensitivity (1000  nm/RIU) and a high Figure of Merit (775 RIU$^{-1}$). Degradation in performance is negligible through  a wide dynamic range up to 0.7 RIU. These quantitative parameters overperform compared to similar plasmonic sensors.

\end{abstract}



0733-8724 \copyright 2019 IEEE. Personal use is permitted, but republication/redistribution requires IEEE permission.
Digital Object Identifier 10.1109/JLT.2019.2906933

\section{Introduction}

Photonic nanostructures  control light propagation through optical media.
They can function as perfect absorbers \cite{liu2010infrared, luo2016perfect}, efficient scatterers 
\cite{ziegler2016plasmonic, jayawardhana2012collection}, frequency selective surfaces 
\cite{monacelli_fss_ieeetap05, debus2007frequency, azemi2018frequency}, etc.
 Optical sensors based on surface plasmon resonances (SPR) benefit from the use of nanostructures for an
increased  range of applications with improved performance. 
For example, they are applied to colorimetry \cite{wong2013colorimetric}, and refractometers for gases \cite{matuschek2018chiral}, bio-fluids \cite{madaan2018ultrahigh}, and chemicals \cite{zhang2018ag}.
A change in refractive index of the media can be measured with a conventional plasmonic device in  
Kretschmann configuration, where the reflectance dip is located and measured angularly \cite{huang2010}. 
In this setup, the key aspects for device performance are: the material and thickness of the metal layer, the refractive index of the prism,  the angle of incidence, and the wavelength of the resonance. 
For angular interrogation, 
the maximum theoretical value of sensitivity of a Kretschmann configuration based sensor is 600 deg/RIU  \cite{huang2010}. 
This extreme value is achieved with a low index prism ($n=1.32$) and an angle of incidence of 
 $ \sim 81^\circ$. 
The need of a low index of refraction is strongly limited by material availability. Some polymers, such as Cytop\texttrademark, reach very low values of the index of refraction ($n_{\rm Cytop}= 1.34$) \cite{zhao2000}. Here, we select magnesium fluoride, MgF$_2$ ($n_{{\rm MgF}_2}= 1.37$), because it is  commercially available, transparent within the spectral region of interest ($\lambda \in [1300, 2000]$ nm), and has a very low index of refraction. Moreover, MgF$_2$  produces planar interfaces when coating nanostructured reliefs \cite{kruger2008}.
The angle of incidence and material constrains of Kretschmann configuration lead to a narrow dynamic detection range \cite{huang2010}. 
To overcome this limit, the community has proposed different configurations of SPR sensors; including spectral interrogation 
and more sophisticated geometries \cite{lee2006gold,shuai2012multi,lan2016highly,luan2017hollow,tong2017optical}.

The  performance parameters of a SPR sensor are: sensitivity, Figure of Merit (FOM), resolution, linearity,  dynamic range, and reproducibility \cite{piliarik2009surface}. 
Sensitivity is the shift of a measurable parameter of the device (a dip in reflectance, transmission, absorption, phase, temperature, etc.) respect to a controlled variable (angle, wavelength, power, etc.) due to a change in the sensed property (refractive index, specimen concentration, color, etc.) \cite{piliarik2009surface}. 
FOM combines sensitivity and the spectral line-shape characteristics in a parameter that compares sensors, independent from the interrogation strategy (angular, spectral, etc.).
Resolution is  the minimum change of the sensed property measured by the device. 
Linearity is assured if sensitivity is constant through the entire dynamic range of the sensor. 
Although high linearity response devices have been reported, the measured shift is not fully  linear  \cite{shuai2012multi}. 
The dynamic range is the interval of the sensed property where the sensor works optimally.  
In refractometric sensors, recent contributions show high efficiency devices with wide dynamic range up to 0.5 RIU \cite{lee2006gold,shuai2012multi,lan2016highly,luan2017hollow,tong2017optical}, including devices with adjustable dynamic range \cite{chen2016refractive}. 
In summary, 
it is very challenging to design an efficient sensor with high sensitivity, large FOM, good resolution, linearity, wide dynamic range, reliability and reproducibility.

 Many approaches enhance performance of SPR optical sensors with bi-metals \cite{goswami2016analysis}, buffer layers \cite{cennamo2017comparison}, and nanoparticles \cite{bolduc2011advances}. 
 The integration of plasmonic sensors with optical fibers and waveguides leads to high performance systems  \cite{ding2017surface} that work in both liquids and gases \cite{gonzalez2018surface}. 
Moreover, other groups show  integrated devices that excite surface plasmon resonances at normal incidence
\cite{elshorbagy2017plasmonic,elshorbagy2017high,
 sun2016integrated,polyakov2012plasmon,lee2010remote,dhawan2011narrow}.
These proposals integrate the sensor at the end of optical fibers or wave-guides. 
Furthermore, the key elements for competitive and practical devices are: feasible and fabricable geometries and features, compatible material combinations for various applications, and high performance.

The operation of a spectrally-interrogated plasmonic sensor relies on the precise measurement of the spectral location of the minimum (or maximum) of its response. 
A narrow spectral peak improves the accuracy of the sensor. In practice,  Fano resonances are an option where the SPR combines coherently with a continuous spectral response \cite{miroshnichenko2010fano}. 
However, this superposition often generates asymmetric spectral line-shapes that can contain both minimum and maximum peaks around a single resonance. 
Also, these duplicated peaks are typically narrower than those obtained for symmetric profiles. 
These spectral features  strongly depend on the characteristics of the coherent superposition responsible for a Fano resonance.
Therefore, the generation of Fano resonances allows to improve  the performance of nanophotonic structures \cite{prodan_hybridizationfano_science2003,somefraud_fanoresonance_2010, gerislioglu_optothermalfano_jlt17, gerislioglu_fanoplasmonic_ieeeptl2017, ahmadivand_fanochemical_jpcc2017} and plasmonic sensors  \cite{yanik_proteinsfano_pnas2017,zhang2018high,deng2018tunable,ai2018strong,
liu2018high,peng2018high,mahboub2018optical,lee2018enhancing}.
Compared to the design presented here, these previous works show lower sensitivities and FOM, and are typically limited by their index of refraction range (table \ref{tab:nonoptimized} in section \ref{sec:characterization} compares the performance parameters of previously reported sensors based on Fano resonances with our design).

Another contribution demonstrates  a SPR sensor that maximizes the electromagnetic energy at the metal/dielectric interface (where surface plasmon are generated)\cite{elshorbagy2017high}. 
The design includes a subwavelength metallic grating to scatter radiation towards the region of interest, while keeping normal incidence. 
In that case, the scattering efficiency  overaffects  the resolution and detection range of the device. In fact, the performance of the device is high for low $n_a$, but decays as $n_a$ increases 
($n_a$ is the index of refraction of the analyte). 
In our design, we enhance the device's characteristics by improving the scattering efficiency and widening the scattering profile. 
Now there is more radiation available to excite resonances even though is normal incidence excitation. 
This is possible with high-contrast index of refraction dielectric subwavelength gratings instead of metallic ones.

The paper is organized as follows. Section \ref{sec:model} describes 
 the Fano resonance parameterizing the SPR, the background spectrum, and the coherent superposition with an analytical model which is applicable to photonic nanostructures.
In subsection \ref{sec:proposedstructure}, we describe the material and geometry of our structure. Then, we analyze the hybrid Fano resonances due to subwavelength dielectric gratings that scatter radiation towards a 
metal/dielectric interface (where the SPR is generated). 
We use COMSOL Multiphysics (COMSOL Inc., Burlington, MA 018013, USA, www.comsol.com) to calculate the spectral response, near-field maps, scattering efficiency, and several relevant device parameters.
We also evaluate and fit the reflectance of the sensor to the model.
Previous contributions show a very good agreement with experimental results and validate this method
\cite{park2015study,yushanov2012surface,athanasopoulos2013enhanced}. 
In subsection \ref{sec:optimization}, our optimization is based on a merit function that combines the quality factor of the resonance with the highest field amplitude delivered at the analyte medium. 
The result is a tunable  Fano-type resonance with very narrow line-width controlled by the geometrical parameters of the grating. 
The optimized sensor is analyzed in section \ref{sec:deviceanalysis}, where we compare
the scattering profile  with the one obtained from optimized metallic gratings  \cite{elshorbagy2017high}. In section \ref{sec:characterization}, we evaluate both the sensitivity and FOM of the proposed sensor, and compare them with previously reported results. For a large dynamic range,
$n \in [1.33, 2.00]$, we have obtained a significant improvement in the sensor performance with a maximum sensitivity of 1000 nm/RIU, and FOM reaching  a maximum of 775 1/RIU. Finally, section \ref{sec:conclusions} summarizes our main findings.
 
 \section{Model, design and optimization}

\label{sec:model}

Fano resonances  manifest  when a discrete state interferes with a continuum of states \cite{miroshnichenko2010fano}.
In our case, the discrete state is given by the excitation of surface plasmons, and the continuous contribution comes from the dispersive character of the subwavelength grating; 
which has a smoother variation than the sharp spectral response of plasmon resonances. 
The coupling between a narrow SPR mode with frequency $\omega_0$ and the broad background spectra creates the simplest scenario for interference and Fano resonances \cite{introoptics1993Fano, fan2002analysis}.
For  a narrow spectral range around the SPR, the slow-varying spectral response of the continuous background, $r_{\rm bk}(\omega)$, can be expanded as a Taylor series without a significant loss in accuracy. 
We can write $r_{\rm bk}(\omega)$ as a polynomial expansion in frequency, $\omega$, centered around the SPR resonance, $\omega_0$:
\begin{equation}
r_{\rm bk}(\omega)= \sum_{j=0}^\infty r_{{\rm bk},j} (\omega - \omega_0)^j
,\label{eq:taylorexpansion}
\end{equation}
where $r_{{\rm bk},j}$ represents the coefficients of the polynomial expansion.
Our model considers a Lorentzian shape for the plasmon response, and a linear variation for the continuous spectrum. 
The linear response is valid as long as the spectral range is narrow enough compared with the width of the SPR and the spectral range considered in the analysis.
Therefore, under the linear approach of the continuous spectral contribution, the spectral reflectivity  can be described as:
\begin{equation}
R(\omega) = \left| \left[ r_{{\rm bk},0} + r_{{\rm bk},1} (\omega - \omega_0 ) \right] + f \frac{\gamma}{(\omega- \omega_0) i + \gamma} \right|^2
.\label{eq:fanomodel_1}
\end{equation}
The term in square brackets is the first order polynomial approximation to the broad spectral response of the system. 
This linearization  works  because the analyzed spectral range, $\Delta \omega$, is quite narrow around the central frequency of the resonance (in our case $\Delta \omega \simeq 0.03 \omega_0$). The second term is a Lorentzian line-shape centered at $\omega_0$, with a width  $\gamma$. 
Both responses are considered in amplitude and are coupled through the complex coefficient $f$, that describes their interference.
We can interpret the coefficients of our model as: 
$r_{{\rm bk},0}$ describes the mean reflectivity for frequencies out of the resonance, and $r_{{\rm bk},1}$ is associated with the slope of this spectral reflectivity. 
The modulus of $f$ describes the interaction of the plasmon resonance with  the broad spectral response. 
If the plasmon resonance produces a dip in reflectance, the phase of $f$ takes values around $\pi$ denoting a minus sign, i.e., a decrease in reflectivity. 
More generally speaking, $f$ is related to the asymmetry of a Fano resonance profile. 
The plasmon resonance is then defined by its central frequency, $\omega_0$, and its width, $\gamma$. 
Finally, a combination of the two defines the quality factor of the resonance: 
\begin{equation}
Q=\omega_0/\gamma,
\label{eq:qualityfactor}
\end{equation}
 that we use in this paper to optimize the analyzed structures.

\subsection{Proposed structure}
\label{sec:proposedstructure}

\begin{figure}[h!]
\centering
  \includegraphics[width=0.80\columnwidth]{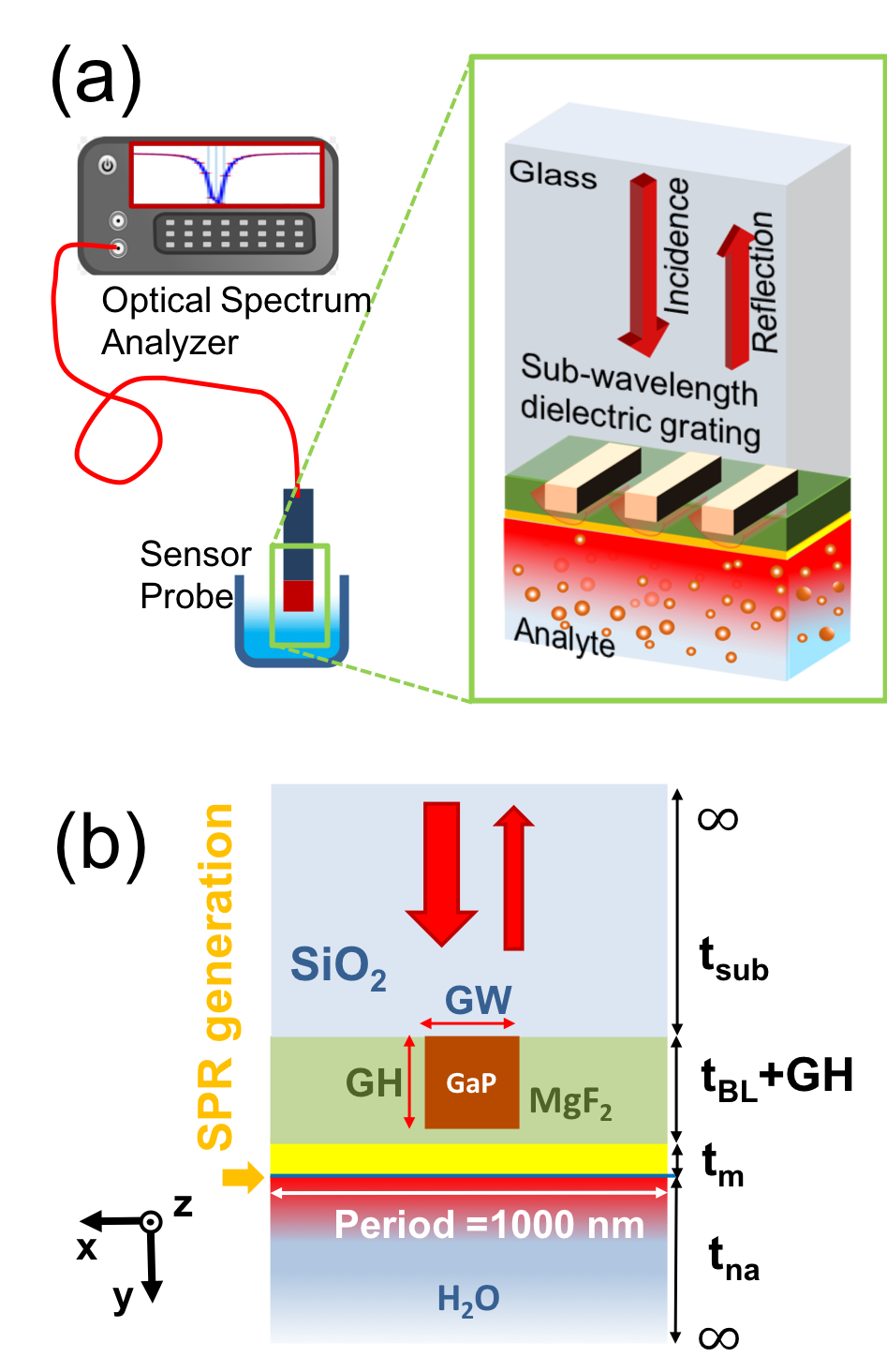}
  \caption{(a) Proposed sketch of our Fano-resonance sensor. The device can be attached to a fiber optics probe  in contact with the analyte. This portion is represented in more detail as a 3D view. (b) Material arrangement, and geometry of the nanostructure of the sensor. Light comes from top to bottom through the substrate (made of SiO$_2$), and reaches the dielectric grating (composed of MgF$_2$ and GaP), and the metallic thin layer (Ag). The SPR takes place at the metal/analyte interface. The analyte is assumed to be liquid (H$_2$O) and it is located at the bottom of this sketch. Silver is selected as the metallic material and it has been represented in yellow. A 10nm-thick passivization layer of MgF$_2$ is added and represented as a thin line located between the metal and the analyte. The reflected power is routed towards an optical spectrum analyzer.}
  \label{fig:design}
\end{figure}

We analyze the Fano resonance in a structure with a 2D periodic dielectric grating  embedded in a dielectric layer. In a practical realization, radiation can be delivered and extracted through a fiber probe (see Fig.~\ref{fig:design}.a).
Figure ~\ref{fig:design}.b shows the material and geometric arrangement of the device, with the substrate located on top  to preserve a typical configuration where the liquid analyte is at the bottom.
The 2D arrangement is extruded along the third dimension ($Z$ axis).
Light arrives under normal incidence conditions from the semi-infinite substrate towards the nanostructure. 
Then, the reflected light is collected by the measurement system, and we monitor the spectral reflectivity. 
  A key point of this proposal is the high contrast in the index of refraction achieved by the combination of two dielectric materials: GaP ($n_{\rm GaP}>3$), and MgF$_2$ ($n_{\rm MgF_2}<1.4$). This arrangement surpasses the limitations of having metallic scatterers within a dielectric matrix \cite{elshorbagy2017high}. 

The device is built on a glass substrate (SiO$_2$). 
The first structure is a 2D grating of GaP rectangles with period $p$, width GW, and thickness (height) GH. The grating relief is spin-coated with MgF$_2$ \cite{kruger2008} that extends above the GaP structure with an additional thickness $t_{\rm BL}$. 
This MgF$_2$ layer generates a plane surface where we deposit an Ag thin layer of thickness $t_m$.
Finally, the structure is coated with an additional MgF$_2$ thin layer that passivizes the properties of silver, maintaining its good optical response and complying with bio-protection standards \cite{sharma2018simulation}.
The sample under analysis, the analyte, is in direct contact with this last surface. 
For biological samples, we select water as the medium in contact with the sensor itself.
The optical constants of the materials involved in this design are taken from recognized sources  
\cite{dodge1984,johnson1972,adachi1989optical}.
The actual structure is formed as:
SiO$_2$ substrate (semi-infinite) / GaP grating (height GH, width GW, period $p$) / MgF$_2$ filling (height GH+$t_{\rm BL}$) / Ag metal film ($t_m$) / MgF$_2$ passivization layer ($t_{\rm PL}=10$ nm) / water analyte (semi-infinite).
At this point,  our results show a negligible dependence on the substrate characteristics; allowing for flexible and plastic substrates without substantial performance degradation.
We can shift the spectral response of the system by varying $p$ (in this paper we take  $p=1000$ nm). 

We analyze our design with   COMSOL Multiphysics to extract information about the spectral reflectivity.
The structure is excited with a plane-parallel monochromatic TM wave (the magnetic field, $H$, has $Z$ component only)  with an input amplitude of 1 A/m. 
 Also, we map the electromagnetic field generated by the structure  to identify the locations where it is enhanced and  where the system sustains plasmon resonances at the desired interface  (device/analyte).

\begin{figure}[h!]
\centering
  \includegraphics[width=0.99\columnwidth]{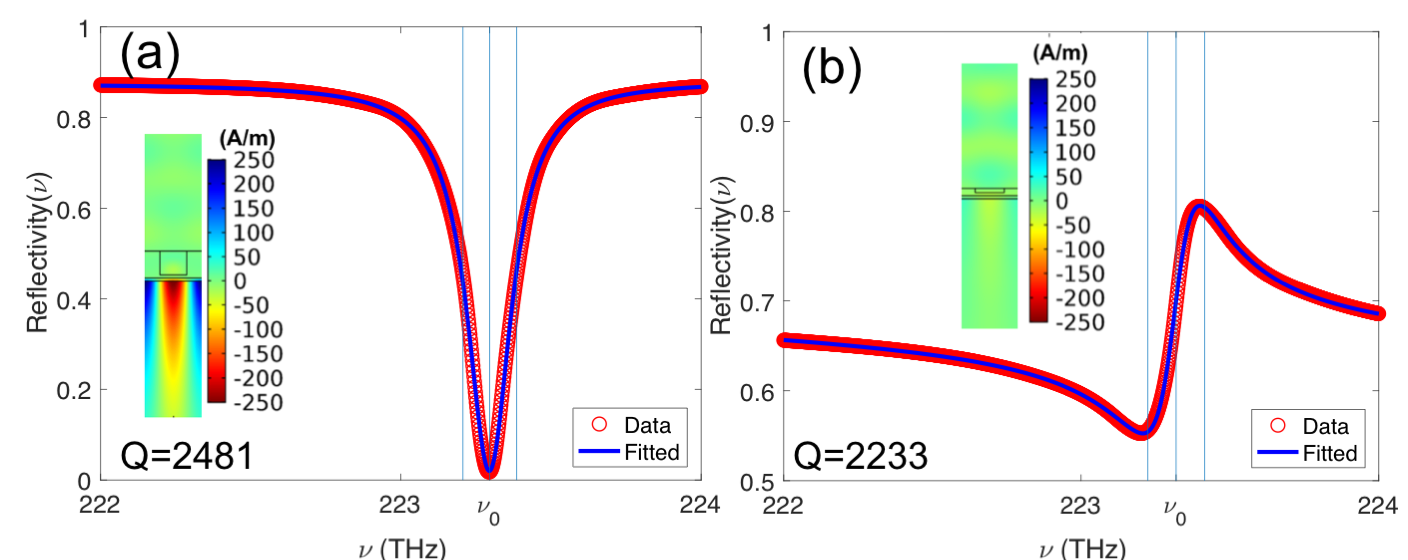}
  \caption{Fitting of computed spectral response (red circles) to the Fano model (blue solid lines) for two different cases showing a symmetric (a) and asymmetric (b) line-shapes. The vertical lines correspond to the central frequency, $\nu_0=\omega_0/2\pi$, and the width given by $\gamma/2\pi$ (expressed in THz) obtained from the model (see Eq. (\ref{eq:fanomodel_1})). The insets represent the near field distributions of the magnetic field, $H_z$, where we can extract the field enhancement parameter. The value of the quality factor of the resonance, $Q=\omega_0/\gamma$, is given for each plot. The fitting coefficients (see Eq. (\ref{eq:fanomodel_1})) for each spectra are: $\nu_0^a=223.30 \rm{ THz}, \gamma^a=0.09 \rm{ THz}, r_{\rm{bk},0}^a=0.94, r_{\rm{bk},1}=0.7\times 10^{-5}$, $f^a=0.795 e^{i 0.99 \pi  }$, and 
  $\nu_0^b=223.32 \rm{ THz}, \gamma^b=0.10 \rm{ THz}, r_{\rm{bk},0}^b=0.81, r_{\rm{bk},1}=1.2\times 10^{-3}$, $f^b=0.156 e^{i 0.48 \pi  }$, where the superscript denotes the plotted spectra (a or b).
  \label{fig:fanofit}
  }
\end{figure}

From a physical point of view, two components combine to generate a Fano line-shape: (i)
 the wide  spectral response generated by the subwavelength dielectric grating, and  (ii) the SPR produced at the metal/dielectric interface. This combination is responsible for the observed Fano resonance that disturbs the SPR and generates symmetric or asymmetric line-shapes depending on their mutual interaction (parameterized with $f$ in Eq. (\ref{eq:fanomodel_1})).
 The subwavelength grating responsible for (i) is made of two dielectric materials (GaP and MgF$_2$) with a high contrast in the index of refraction. 
Figures \ref{fig:fanofit}.a and \ref{fig:fanofit}.b show how the numerically evaluated reflectivity fits  with the model (Eq. (\ref{eq:fanomodel_1})) for two cases: (a) symmetric line-shape with an almost pure Lorentzian shape, and (b) asymmetric.
The fit provides the  parameters of the model that quantifies the design. Among these parameters, we chose  $\omega_0$ and $\gamma$ to characterize the performance of the resonance for sensing applications through the definition of the quality factor of the resonance, $Q$ (see Eq. (\ref{eq:qualityfactor})).  
As the geometrical parameters vary, the coupling between this two contributions changes too, resulting in symmetric (Fig. \ref{fig:fanofit}.a)  or asymmetric (Fig. \ref{fig:fanofit}.b) line-shapes. The symmetric case (a) corresponds with an almost real value of $f^a=0.795 e^{i 0.99 \pi  }$. The asymmetric case (b) shows an almost imaginary value of $f^b=0.156 e^{i 0.48 \pi  }$. 
Both cases are well described through the hybrid plasmonic-Fano resonance model in Eq. (\ref{eq:fanomodel_1}).

 Figure \ref{fig:fanofit} includes insets of the  maps of the magnetic field amplitude, $H_z$, that shows the field enhancement located at the region of interest, where the SPR is excited and propagates through the analyte within a given interaction volume at the metal/analyte interface (actually, the analyte is in contact with the MgF$_2$ passivization layer). From the magnetic field maps, we define the field enhancement parameter, FE, that describes the maximum field within the interaction volume for a given geometry of the device. Actually, FE is the ratio between the absolute value of the field and the amplitude of the incoming wavefront (we have used an input amplitude of 1 A/m) and therefore, the absolute value of the magnetic field (in A/m) is equal to FE. 
The interaction volume comprises the region in the analyte where the plasmon resonance propagates. 
Later on, in section \ref{sec:deviceanalysis} we define a quantitative parameter, $\delta_y$, that can be used to evaluate this volume. 

\subsection{Optimization}
\label{sec:optimization}

\begin{figure}[h]
\centering
  \includegraphics[width=0.80\columnwidth]{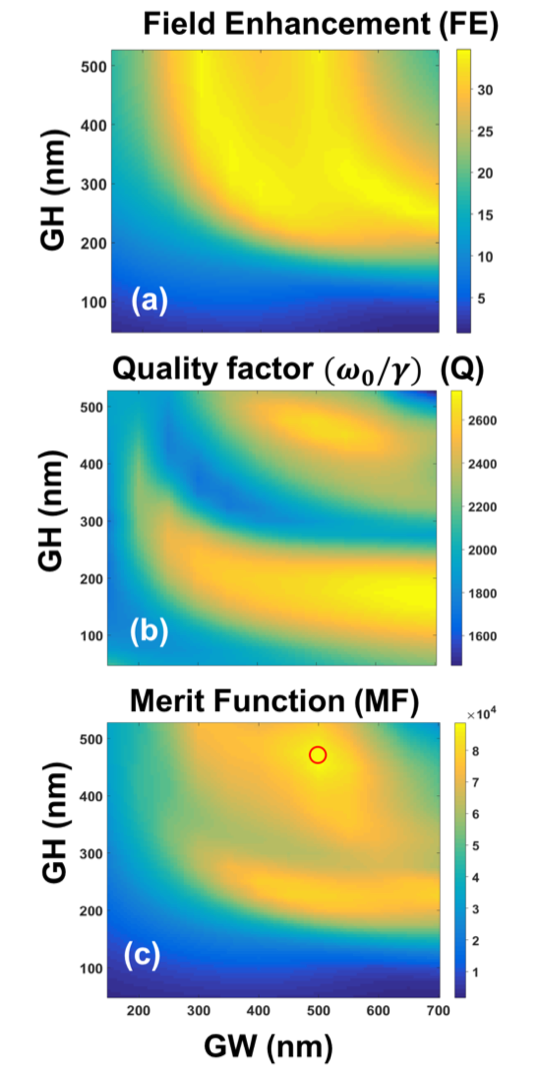}
  \caption{Maps of the quality factor (a), $Q$, the field enhancement parameter (b), FE, and the Merit Function (c) defined in  Eq. (\ref{eq:merit}), as a function of the geometrical parameters GW and GH. These maps are used to select the values for an optimum performance of the system. The red circle in the Merit Function map corresponds with its maximum value obtained at GH=475 nm, and GW=500 nm.}
  \label{fig:model}
\end{figure}

We optimize the performance of the device as a plasmonic sensor
by combining two parameters that describe both the sharpness of the resonance, and the amount of light that interacts with the analyte. These two parameters are the quality factor of the resonance, $Q$ (see Eq. (\ref{eq:qualityfactor})),
 calculated from the fitting of the spectral results with the model (see 
Eq. (\ref{eq:fanomodel_1})), and the field enhancement values, FE, obtained at the region of interaction. They are both included into a Merit Function as the geometrical mean: 
\begin{equation}
\mbox{Merit Function}= \sqrt{ Q \times {\rm FE} }
.\label{eq:merit}
\end{equation}
We choose this merit function after evaluating other combinations of these two parameters, including the arithmetic mean, and the root mean squared. All of them produced the same output for the geometric parameters (GW and GH) of the structure where the Merit Function is maximized.
We present the results of the optimization process for different parameters in Fig. \ref{fig:model}.
 
The maps in Figs. \ref{fig:model}.a and \ref{fig:model}.b represent the values of the quality factor, $Q$, and the field enhancement parameter, FE, in terms of two geometric parameters of the dielectric grating: GW, and GH. 
We choose first these parameters as optimization variables because  they are the main players in the process, as shown in previous results \cite{elshorbagy2017high, elshorbagy2017plasmonic}. 
This is why the rest of the parameters, including the grating period $p$, 
are kept constant.  
Figure \ref{fig:model}.c  contains the Merit Function map that identifies a region of values that combines a high value of $Q$ and FE. 
Our evaluation provides a value of GW=500 nm and GH=475 nm (presented in Fig. \ref{fig:model}.c as a red circle). 
This optimization ignores the symmetry of the resonance, as far as both (symmetric and asymmetric) line-shapes are well described through the plasmonic-Fano resonance (Eq. (\ref{eq:fanomodel_1})). 
We have checked that the resonance is almost symmetric at the optimum point in Fig. \ref{fig:model}.c.

\begin{figure}[h]
\centering
  \includegraphics[width=0.99\columnwidth]{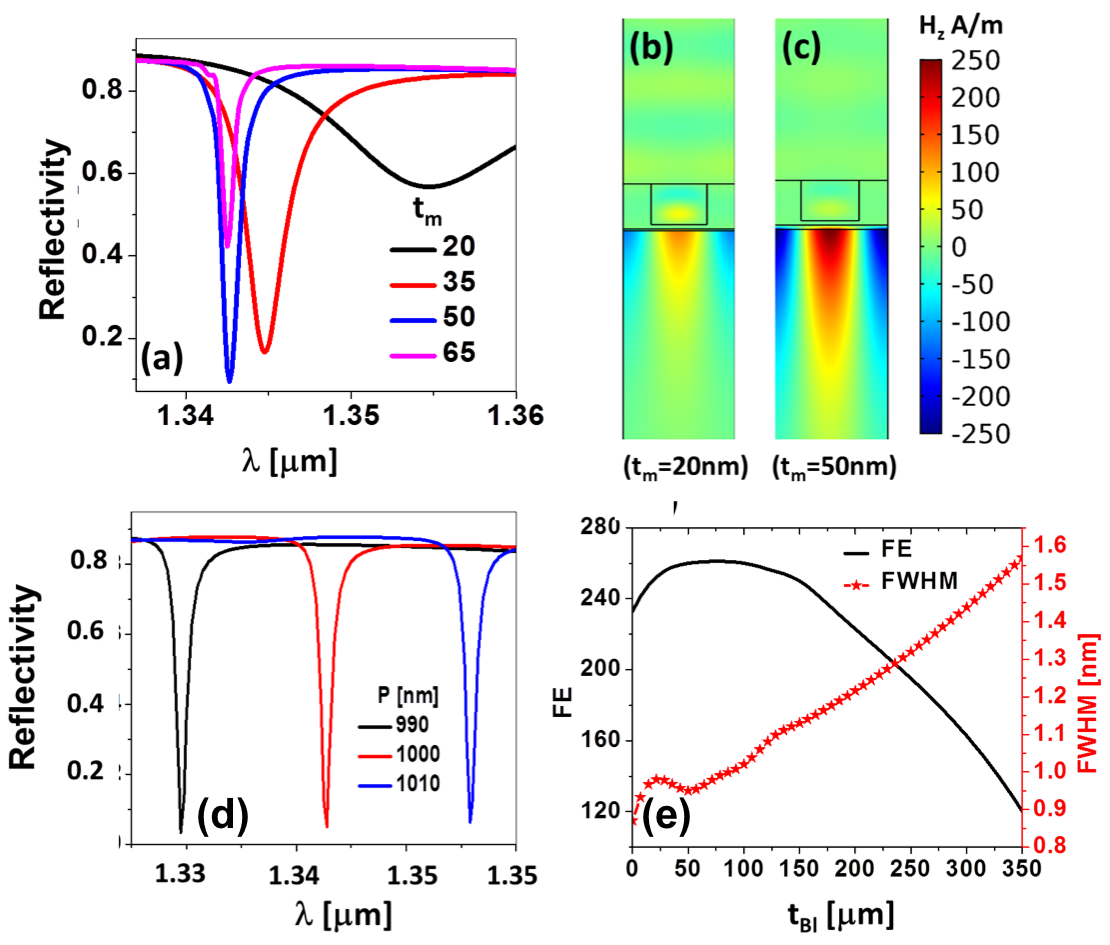}
  \caption{(a) Spectral reflectivity for several values of the metal layer thickness, 
$t_m$. (b) and (c) Near field amplitude maps showing the field enhancement for two values of
$t_m=20$ nm, and $t_m=50$ nm, where the last value clearly provides a larger amplitude within the interaction volume. (d) Spectral reflectivity for three values of the period of the grating, $p$. (e) Variation of the field enhancement parameter (FE), and the FWHM of the resonance as a function of the thickness of the buffer layer, $t_{\rm BL}$.}
  \label{fig:thickness}
\end{figure}
 
Once the shape of the grating is fixed, we analyze other geometrical parameters. The thickness of the metallic layer, $t_m$, affects  the
 surface plasmon generated at the metal dielectric interface and its longitudinal range. 
 Through further optimization, we find a  value of $t_m=50$ nm. 
This evaluation is represented in Fig. \ref{fig:thickness}.a where the spectral shape allows to choose $t_m=50$ nm as the one showing the smallest line-width in reflectance, and a larger field enhancement.
The maps of the magnetic field are plotted in Fig. \ref{fig:thickness}.b and .c for two cases ($t_m=20$ nm and $t_m=50 $ nm). We see that the field is stronger for $t_m=50$ nm, that is the thickness chosen for the metallic layer. As far as the input wavefront has an amplitude of 1 A/m, the absolute value of the plotted amplitude of the magnetic field is FE.
We also analyzed the period of the grating, $p$, and its effect is shown in Fig. \ref{fig:thickness}.d. The spectral reflectivity shifts towards longer wavelengths when the period increases. 
Also, through another analysis to optimize the thickness of the MgF$_2$, $t_{\rm BL}$, we found a small performance change for values between 10 and 100 nm. We found that  
  $t_{\rm BL}=50$ nm  produces a minimum width (minimum FWHM), and a maximum FE (see Fig. \ref{fig:thickness}.e).
 Also, we set the passivization layer at 
$t_{\rm PL}=10$ nm (minimum fabricable thickness) to allow the best plasmon propagation possible.

\section{Device analysis}
\label{sec:deviceanalysis}

In this section, we discuss the device's performance compared with alternative designs, including the classical Kretschmann configuration.

\begin{figure}[h]
\centering
  \includegraphics[width=0.80\columnwidth]{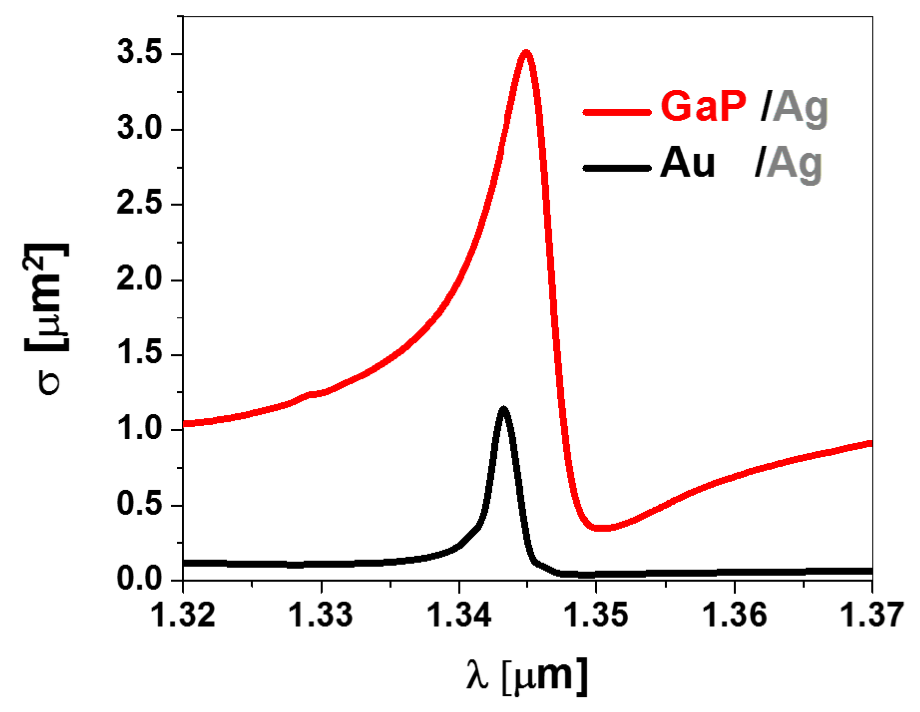}
  \caption{Scattering cross section, $\sigma$,  for the proposed geometry with GaP grating (red solid line) and Au grating (black solid line) 
  \cite{elshorbagy2017high}. Ag is used at the metal/analyte interface.}
  \label{fig:scatter}
\end{figure}

A previous contribution  optimized the material setup of a subwavelength metallic grating 
to maximize the amount of energy transferred to the metal surface for  SPR generation \cite{elshorbagy2017high}. 
In the design presented here, 
a subwavelength dielectric grating, involving a nanostructure with a high-contrast of the index of refraction, scatters radiation and reveals very narrow widths of Fano-type resonances, and optimum coupling to the metallic/analyte interface. 
This provides a  higher scattering efficiency (see Fig.  \ref{fig:scatter}).  
Both metallic and dielectric gratings create a continuum diffraction pattern (described at the first term in Eq. (\ref{eq:fanomodel_1})).  
Unfortunately, 
 metallic gratings support modes where the incoming wave is absorbed, scattered, reflected,  or adds to SPR on the grating surface itself. 
On the contrary, The dielectric GaP grating are transparent at the operating wavelengths. 
Moreover, it only generates guiding modes at its surface, and avoids the excitation of additional SPRs. 
To understand the advantages of our GaP structure  with respect to a metallic subwavelength element, we  compare the enhancement of the spectral characteristics of both gratings.  
Figure \ref{fig:scatter}
represents the scattering cross-section of these subwavelength gratings, and shows that for dielectric materials there is 3 times larger scattering cross-section at the resonant wavelength. Therefore, more radiation is scattered towards the metal/dielectric interface where the SPR is generated. 
In contrast, the metal grating absorbs radiation and redirects part of it towards modes appearing at the unavailable inner metal/dielectric interface \cite{elshorbagy2017high}. 
In both cases, the geometries were optimized for optimum performance.

\begin{figure}[h]
\centering
  \includegraphics[width=0.99\columnwidth]{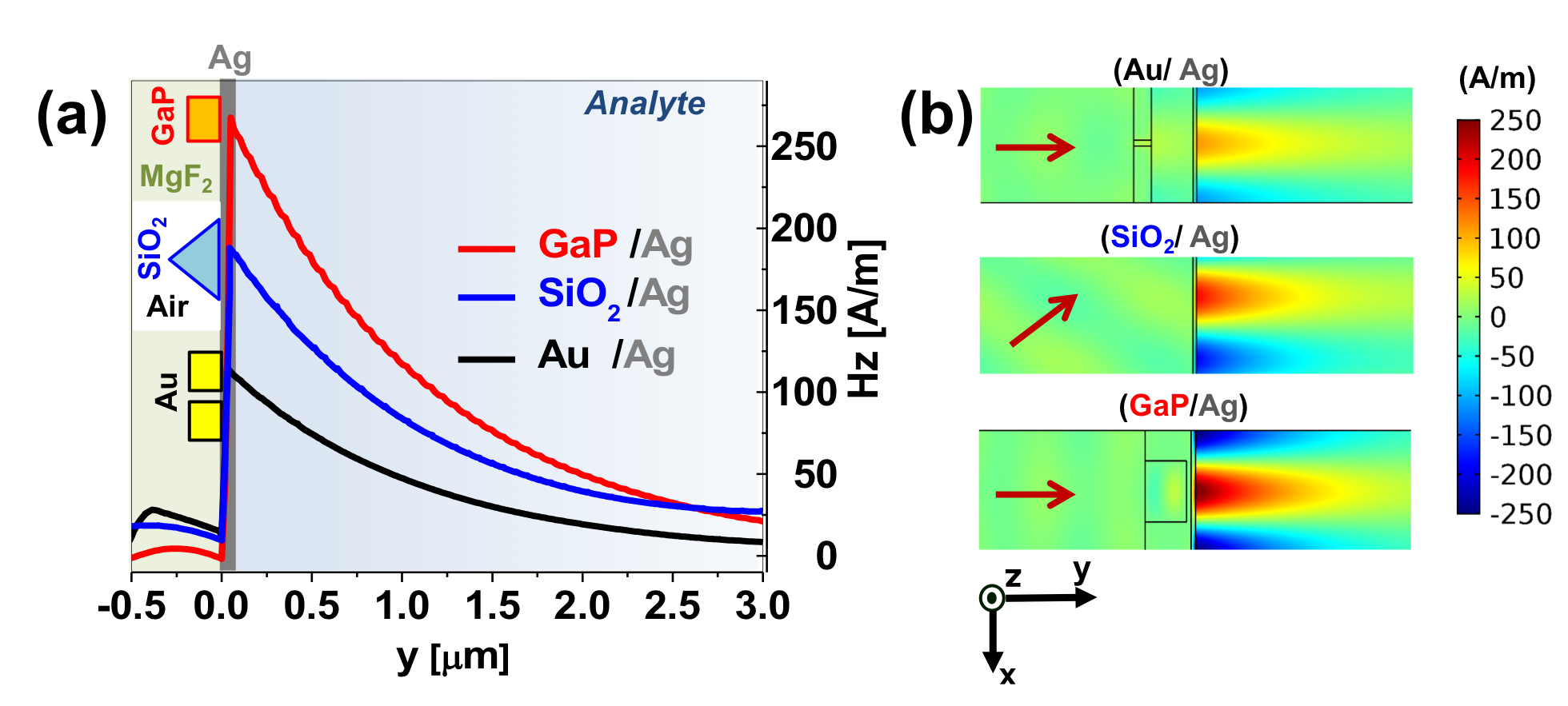}
  \caption{(a) Evolution of the near field in the region of interest for three different geometries. GaP/Ag denotes the design presented in this paper. SiO$_2$/Ag corresponds to the classical Kretschmann configuration. Au/Ag shows the response for a subwavelength metallic grating. (b) Magnetic field amplitude distributions for the three cases.}
  \label{fig:response}
\end{figure}

The interaction volume describes the portion in the analyte that is exposed to a large value of electromagnetic field due to the generation and propagation of the plasmon resonance. This interaction volume increases with an optimized geometry that gives a large value of FE.
In figure \ref{fig:response}.a,
we compare the amplitude of the magnetic field, $H_z$, within the analyte for three  cases: (i) our design, (ii)  Kretschmann configuration, and (iii) metallic subwavelength grating structure. 
All these three designs were optimized for later comparison. The parameters of our design (i) have been obtained through the method presented in section \ref{sec:optimization}, the Kretschmann configuration (ii) corresponds with an optimum choice of materials and angle of incidence (as described in the introduction section), and the metallic subwavelength grating (iii) was derived from previous published results \cite{elshorbagy2017high}.
We fit the amplitude dependence with distance to an exponential decay function \cite{elshorbagy2017high}: 
\begin{equation}
H_z = H_{z,0} \exp\left( -\frac{y}{\delta_y} \right) 
,\label{eq:expondecay}
\end{equation}
where $H_{z,0}$ is the maximum amplitude obtained at the metal/analyte interface.
The enhancement factor amplitude of the maximum field is $\times 277$, $\times 183$, and $\times 116$ for our structure (i), the Kretschmann setup (ii), and metallic grating case (iii), respectively.
Figure \ref{fig:response}.b represents the magnetic field for the three cases.
 This translates in a $\times 1.5$ larger   field with our system compared with the Kretschmann configuration, and more than double respect to the metallic grating case.
Furthermore, the decay lengths, $\delta_y$, are 1162 nm, 1013 nm and 1085 nm for our structure (i), the Kretschmann setup (ii), and  metallic grating case (iii), respectively.
These numbers are derived after fitting the amplitude evolution presented in Fig. \ref{fig:response}.a with the model in Eq. (\ref{eq:expondecay}). As far as the input amplitude is normalized at 1 A/m, $H_{z,0}$ represents the field enhancement.
Therefore, our design generates  a larger volume of interaction, and has a potentially lower limit of detection.
These results surpass those obtained for metallic grating structures, and they are significantly better than classical Kretschmann configuration.

\section{Characterization of the hybrid plasmonic-Fano resonance sensor}

\label{sec:characterization}

\begin{figure}[h!]
\centering
  \includegraphics[width=0.80\columnwidth]{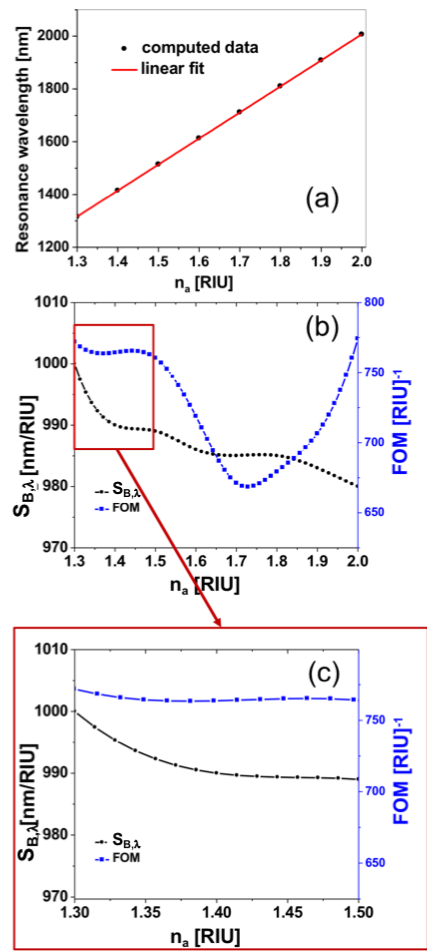}
  \caption{(a)  Dependence of the central frequency of the resonance as a function of the index of refraction of the analyte. The red line corresponds with the linear fitting of the numerical results. (b) Sensitivity, $S_{B,\lambda}$, and FOM as a function of the index of refraction of the analyte, $n_a$. (c) Detailed portion of (b) showing a range in the refractive index that maintains an almost constant behavior of the device.}
  \label{fig:sens}
\end{figure}

Until now,  high-sensitivity plasmonic sensors operating under normal incidence and based on spectral measurements show important limitations in detection range and linearity \cite{elshorbagy2017plasmonic,elshorbagy2017high,sun2016integrated,
polyakov2012plasmon,dhawan2011narrow}. 
These limitations are related with the presence of unprofitable modes where the  field enhancement is not localized at the metal/analyte interface \cite{elshorbagy2017high}. 
Our device is linear in a wide range of index of refraction of the analyte, $n_a \in [1.3, 2]$,
thanks to the high coupling efficiency of the dielectric subwavelength grating,
which forwards the radiation  into the interaction interface. 
Figure \ref{fig:sens}.a shows the linear dependence of the  SPR resonant wavelength  with the index of refraction of the analyte, $n_a$.  
This linear behavior is linked with the smooth dependence of $n_{\rm GaP}$ and $n_{{\rm MgF}_2}$, that do not present absorption peaks along the analyzed wavelength range. 
This broadband feature permits multifunctional operation  in biological applications in water, glucose, urea,  etc.\cite{verma2014novel}.

To validate our proposal as a refractometric sensor, we evaluate  sensitivity, $S_{B,\lambda}$, and Figure of Merit, FOM, which are defined as \cite{piliarik2009surface, huang2010}:
\begin{equation}
S_{B,\lambda}=\frac{\Delta \lambda_M}{\Delta n}
,\label{eq:sensitivity}
\end{equation}
 and
\begin{equation}
{\rm FOM}= \frac{S_{B,\lambda}}{{\rm FWHM}}
\label{eq:figureofmerit}
,\end{equation}
where $\lambda_M$ denotes the location of the minimum (or maximum) of the reflectivity (transmisivity) of the retrieved spectrum, and FWHM refers to the full-width-at-half-maximum (width of the spectral feature). 
For a system that performs with a maximized Merit Function  
(see Eq. (\ref{eq:merit}) in subsection \ref{sec:optimization}), we evaluate the shape and characteristics of the spectral resonance line-shapes
through  the width of the Lorentzian term, $\gamma$, of the Fano resonance (see Eq. (\ref{eq:fanomodel_1})), and identify it  with the FWMH appearing in Eq. 
(\ref{eq:figureofmerit}). 
To do so, we fit the numerically evaluated spectral response with the analytical model 
presented in section \ref{sec:model}  (see Eq. (\ref{eq:fanomodel_1})).
This eliminates any ambiguity due to asymmetric profiles, and makes the procedure robust. 

\begin{table*}[h!]
\begin{center}
\caption{Comparison of sensor performance versus the results obtained in this work. All the sensors are spectrally interrogated and are based on Fano resonances. 
  \label{tab:nonoptimized}
  }
\resizebox{\textwidth}{!}{%
\begin{tabular}{ccccc}
\hline \hline
Structure &  Sensitivity &  FOM &  Dynamic Range & Ref 
\\  \hline \hline
periodical asymmetric paired bars  & 370 [nm/RIU] & 2846 [1/RIU]&1.333-1.334 & \cite{zhang2018high}
\\ \hline
metal-dielectric-metal waveguide coupled to a pair of cavities  & $1.1\times10^{8}$ [1/m] & $2.33\times10^{4}$ [1/RIU] & 1.00-1.04 & \cite{deng2018tunable}
\\ \hline
array of nanoparticle-in-ring nanostructures  & 425 [nm/RIU] & 4.7-7.1 [1/RIU]&1.33-1.50 & \cite{ai2018strong}
\\ \hline
split-ring metasurface &  452  [nm/RIU] & 56.5 [1/RIU] &1.0-1.2& \cite{liu2018high}
\\ \hline
photonic crystal cavity-coupled microring resonator &  67 [nm/RIU] & ---&1.332-1.344 & \cite{peng2018high}
\\ \hline
MIM waveguides with a double split ring resonator & 586 [nm/RIU]  & 62 [1/RIU] &1.00-1.05 & \cite{mahboub2018optical}
\\ \hline
periodic silver nanoslit array coupled to a cavity&  900 [nm/RIU] & ---- &1.33-1.36& \cite{lee2018enhancing}
\\ \hline
diffraction grating coupled to thin metal film &  1000 [nm/RIU] & 775 [1/RIU] &1.33-2.00 & this work
\\ \hline \hline
\end{tabular}}
\end{center}
\end{table*}

From the FWHM, we  obtain the values of sensitivity and FOM for the optimized device (see 
Fig. \ref{fig:sens}.b). The maximum sensitivity is $S_{B} \approx 1000$ nm/RIU, and the maximum FOM is $\approx$ 775 RIU$^{-1}$. 
These values are higher than recently reported values
for similar sensors \cite{ding2017surface,wu2017,meng2017figure,pandey2018simulation,zhang2018plasmonic,rahman2018novel,
xiang2018highly,luan2019enhanced}. 
Fiber optics based sensors exhibit higher sensitivity than our proposal, but present a lower FOM  \cite{rifat2018highly,sharma2018review}.   
Experimentally \cite{liu2018plasmonic}, scientists   achieve a FOM of 730 RIU$^{-1}$ that leads to detection of ultra-low concentrations of $10^{-10}$ M. 
In figure \ref{fig:sens}.b the sensitivity of our design drops from 1000 to 980.5 nm/RIU   at $n_{a} = 2$,  and the FOM has a minimum value of 667 RIU$^{-1}$ at  $n_a=$ 1.725.
These figures are competitive for our design because it extends the range of operation to higher
index of refraction if compared to 
refractomeric sensor based on Fano resonances (see Table \ref{tab:nonoptimized}). 
Moreover, the FOM must be considered to calculate the device resolution, or more specifically, the spectral width.  Sensitivity and FOM are almost constant in the range 
$n_a \in [1.3, 1.5]$, as shown in Fig.  \ref{fig:sens}.c.
This constant trend indicates  high stability and reliability in the measurements, 
resulting in a high performance device.

Table \ref{tab:nonoptimized} summarizes the performance of several sensors with different methods
involving Fano resonances. As we mention in the introduction section, a high sensitivity and figure of merit over a wide range of refractive index are key for multifunctional and efficient sensors.  
However, in most cases, enhancing one of these parameters diminishes the others. Sometimes, the good performance of a given
sensor is strongly degraded when moving away from the optimum point of operation of 
the value of the analyte refractive index \cite{deng2018tunable}. 
Our device performs better in sensitivity (Eq. (\ref{eq:sensitivity})), figure of merit (Eq. (\ref{eq:figureofmerit})), and dynamic range (in terms of variation of the index of refraction of the analyte). The sensor is interrogated spectrally. However, other  methods, such as amplitude interrogation, may provide sensitivity values of one order of magnitude larger than those reported for spectral interrogation \cite{qiu2016plasmonic}.

\section{Conclusions}
\label{sec:conclusions}

The sensing community has defined
sensitivity and Figure of Merit as quality parameters to compare refractometric sensors. 
These performance parameters can be evaluated from the spectral  line-shape of the resonance
when the system monitors and measures the change in the spectral response. 
A hybrid plasmonic-Fano resonance can be created when a narrow SPR  coherently combines with a continuous spectrum. 
Moreover, a Fano resonance may show a narrower spectral response along with an asymmetric behavior,
making it a good  candidate for improved performance plasmonic sensors. 
In this contribution, we show a physical model of the Fano resonance, and test it through simulation. 
The model superimposes  a smooth spectral variation (approximated as a linear spectrum) with a Lorentzian shape (characteristic of a  SPR).
We have interpreted and related the parameters of our model with the elements that define both sensitivity and Figure of Merit. 
Furthermore, we parameterize the Lorentzian contribution  with the central frequency, $\omega_0$, and width,  $\gamma$; which we then use to evaluate  $S_{B,\lambda}$, and FOM (after the corresponding transformation into wavelengths).

Based on previous results, we propose a  simple plasmonic device that generates Fano resonances using a dielectric subwavelength grating. 
The resonance occurs at the metal/analyte interface due to the scattering originated from a high-contrast index of refraction structure modeled as a 2D extruded grating. We chose dielectric materials to reduce absorption and avoid  ineffective  
resonances that propagate as guiding modes in metallic structures. 
This structure operates at normal incidence  and outperforms  previously reported  metallic subwavelength gratings by doubling the field enhancement at the region of interest while FOM is also increased by 26\%. 

We optimized the device adjusting its  geometrical parameters: shape of the grating,  thicknesses of the metallic layer supporting the SPR, and the planarization dielectric layer. 
Moreover, the design includes a 10 nm-thick passivization layer to exploit the
plasmonic response of silver and preserve the analyte biocompatibility. 
To optimize the sensor, we have defined a Merit Function that combines the quality factor of the SPR and the
field enhancement  at the measurement interface. 
Afterwards, we numerically tested the proposed device to  calculate both the sensitivity and the Figure of Merit of the sensor. 
From our results, our design has the following characteristics: 
(i) operates linearly for $n \in [1.3, 2]$; 
(ii) sensitivity remains between 980 and 1000 nm/RIU,
and (iii) FOM varies between 667 and 775 RIU$^{-1}$. 

In summary, our design uses hybrid plasmonic-Fano resonances to improve the performance of refractometric sensors. 
This  device can be integrated at the end of an optical fiber and also works under normal incidence illumination. 
It takes advantage of the  non-absorbing properties of the high-contrast index of refraction subwavelength dielectric gratings.  
Also, the geometry and dimensions consider fabrication constraints. 
It provides a  large 
sensitivity and figure of merit within a wide range in the index of refraction of the analyte, thus suitable for multifunctional sensing. 
This makes our system very competitive compared to previously reported results.

\section*{Acknowledgements}
This work has been partially supported by Ministerio de Econom\'{\i}a y Competitividad of Spain (MINECO) (TEC2013-40442), and by Ministry of Higher Education of Egypt (MOHE) (missions section). 
F. J. Gonz\'{a}lez would like to acknowledge support from Project 278291 (SRE-CONACYT), project 105 of ``Centro Mexicano de Innovaci\'{o}n en Energ\'{\i}a Solar"
and by the National Laboratory Program from Consejo Nacional de Ciencia y Tecnolog\'{\i}a of Mexico (CONACYT) through the Terahertz Science and Technology National Lab (LANCYTT).
The authors thank Irene Alda for her critical and careful reading of the manuscript, and the fruitful discussion along the development of this contribution.

\bibliographystyle{IEEEtran}
\bibliography{SGS} 

\end{document}